\documentstyle[11pt,aaspp4]{article}

\def\hst{{\em HST}}
\def\hnot{H$_0$}
\def\qnot{q$_0$}
\def\msun{\ifmmode {\rm M_\odot} \else M$_\odot$\fi}
\def\lsun{\ifmmode {\rm L_\odot} \else L$_\odot$\fi}

\def\kms{km s$^{-1}$}
\def\deg{\ifmmode ^{\circ}
         \else $^{\circ}$\fi}
\def\pdeg{\ifmmode 
           $\setbox0=\hbox{$^{\circ}$}\rlap{\hskip.11\wd0 .}$^{\circ}
     \else \setbox0=\hbox{$^{\circ}$}\rlap{\hskip.11\wd0 .}$^{\circ}$\fi}

\def\arcsec{\ifmmode '' \else $''$\fi}
\def\arcsecpt{\ifmmode ''\!\!. \else $''\!\!.$\fi}

\def\msunyr{\ifmmode {\rm M_\odot~yr^{-1}}\else${\rm M_\odot~yr^{-1}}$\fi}
\def\lam{\ifmmode {\lambda} \else {$\lambda$} \fi}

\def\mdoto{\ifmmode {\dot{M}_0} \else  $\dot{M}_0$ \fi}
\def\teff{\ifmmode {T_{eff}} \else $T_{eff}$ \fi}
\def\ilam{\ifmmode {I_\lambda} \else  $I_\lambda$ \fi}
\def\inu{\ifmmode {I_\nu} \else  $I_\nu$ \fi}
\def\fnu{\ifmmode {F_\nu} \else  $F_\nu$ \fi}

\def\yr{\ifmmode {\rm yr} \else  yr \fi}
\def\cm{\ifmmode {\rm cm} \else  cm \fi}
\def\cmmitwo{\ifmmode \rm cm^{-2} \else $\rm cm^{-2}$\fi}
\def\cmmithree{\ifmmode \rm cm^{-3} \else $\rm cm^{-3}$\fi}
\def\cmps{\ifmmode \rm cm~s^{-1}\else $\rm cm~s^{-1}$\fi}
\def\cmpsps{\ifmmode \rm cm~s^{-2}\else $\rm cm~s^{-2}$\fi}
\def\kmps{\ifmmode \rm km~s^{-1}\else $\rm km~s^{-1}$\fi}
\def\kmpspmpc{\ifmmode \rm km~s^{-1}~Mpc^{-1} \else
    $\rm km~s^{-1}~Mpc^{-1}$\fi}
\def\ergps{\ifmmode \rm erg~s^{-1} \else $\rm erg~s^{-1}$ \fi}
\def\ergpspcm{\ifmmode \rm erg~s^{-1}~cm^{-2} \else $\rm erg~s^{-1}~cm^{-2}$ \fi}
\def\ergpspcmphz{\ifmmode \rm erg~s^{-1}~cm^{-2}~Hz^{-1} \else $\rm
erg~s^{-1}~cm^{-2}~Hz^{-1}$ \fi}
\def\ergpspcmpa{\ifmmode \rm erg~s^{-1}~cm^{-2}~\AA^{-1} \else $\rm
erg~s^{-1}~cm^{-2}~\AA^{-1}$ \fi}
\def\ergpsphz{\ifmmode \rm erg s^{-1} Hz^{-1} \else 
   $\rm erg s^{-1} Hz^{-1}$ \fi} 

\def\eg{e.g.}

\def\etal{~et al.}

\received{}
\accepted{}
\journalid{}{}
\articleid{}{}

\slugcomment{Accepted for Publication in Astrophysical Journal Letters
\\Volume 550 (2001 March 20)}

\begin{document}

\title{LBQS 0103$-$2753: A 0.3 Arcsec Binary Quasar\altaffilmark{1}}

\author{V. Junkkarinen\altaffilmark{2,3},
G. A. Shields\altaffilmark{4},
E. A. Beaver\altaffilmark{2},
E. M. Burbidge\altaffilmark{2},
R. D. Cohen\altaffilmark{2},
F. Hamann\altaffilmark{5},
R. W. Lyons\altaffilmark{2}
}

\altaffiltext{1}{Based on observations made with the NASA/ESA Hubble
Space Telescope.  STScI is operated by the Association of Universities
for Research in Astronomy, Inc. under NASA contract NAS5-26555. }

\altaffiltext{2}{Center for Astrophysics and Space Sciences,
University of California, San Diego, La Jolla, CA 92093-0424; vesa@ucsd.edu,
mburbidge@ucsd.edu, rdcohen@ucsd.edu.}

\altaffiltext{3}{Visiting Astronomer, Cerro Tololo Inter--American
Observatory, which is operated by the Association of Universities
for Research in Astronomy, Inc., under contract with the National
Science Foundation.}

\altaffiltext{4}{Department of Astronomy, University of Texas, Austin TX
78712; shields@bluebump.as.utexas.edu.}

\altaffiltext{5}{Department of Astronomy, University of Florida, Gainsville FL 32611-2055; hamann@astro.ufl.edu.}

\lefthead{V. Junkkarinen}
\righthead{Binary QSO}


 
\begin{abstract}

Imaging and spectroscopy with \hst\ show that LBQS 0103$-$2753 
(V = 17.8, $z$ = 0.848)
is a binary quasar with
a separation of 0\arcsecpt3  or 2.3 kpc.  
This is by far the smallest separation 
binary quasar reported to date.
The two components have very
different spectra, including the presence of 
strong broad absorption lines (BALs) in
component A only.
The emission-line redshifts, based on the broad high
ionization C IV lines, are $z_A$ = 0.834 and $z_B$ = 0.858; 
their difference is 3900 \kms\ in velocity units. 
The broad C IV lines, however, are probably not
a good indicator of systemic redshift; and LBQS 0103$-$2753 A and B could
have a much smaller systemic redshift difference, like the other known
binary quasars. If the systemic redshift difference is small, then LBQS
0103$-$2753 would most likely be a galaxy merger that has led to a binary
supermassive black hole.
There is now one known 0\arcsecpt3 binary among roughly 500 QSOs that have been
observed in a way that would reveal such a close binary. This suggests that QSO
activity is substantially
more likely for black hole binaries at spacings $\sim2$ kpc than at
$~\sim15$ to 60 kpc.  Between 1987 and 1998, the observed Mg II BAL disappeared.

\end{abstract}


\section{INTRODUCTION}

Quasar pairs are known to represent three
classes of objects: chance alignments, gravitational
lenses, and binary quasars.
Binary quasars are rare; in the LBQS sample of 
$\sim$10$^3$ quasars, Hewett et al. (1998) find two
binaries.  A similar rate of occurrence is found for
all quasars in the literature: 11 QSO pairs
in $\sim$ 10$^4$ quasars (Kochanek, Falco,
and Mu\~noz 1999).  This rate is high
compared to a naive extrapolation of the quasar--quasar
two-point correlation function at megaparsec
separations to separations less than 100 kpc (Djorgovski 1991).
The existence of binary quasars, as
distinct from the gravitationally lensed pairs,
has been established
by the occurrence of quasar pairs with only one
radio loud member (e.g. Djorgovski et al. 1987) and
by the statistics of QSO pairs with large ($\ge$ 3\arcsecpt0)
separations (Kochanek \etal\ 1999).
Kochanek et al. find that quasar pairs with
$\Delta$$\theta$ $\le$ 3\arcsecpt0 are gravitational lenses
and that most quasar pairs with
$\Delta$$\theta$ $>$ 3\arcsecpt0 are binary quasars.
The leading explanation for binary quasars is the fueling
of quasar activity when two galaxies with massive black holes
in their nuclei collide (Kochanek \etal\ 1999).

The observation that two nearby quasar images have similar
emission-line spectra is insufficient to allow
classification as a lens (e.g. Mortlock, Webster, and Francis 1999); and
gravitational lens pairs can have different emission-line
spectra due to reddening or time delays.
However, quasar pairs with almost identical redshifts but very
different emission-line spectra or BALs are probably binary quasars.
The cloverleaf QSO, H1413+114, with $\Delta\theta_{max} = 1\arcsecpt4$,
is a four image
gravitational lens; all four images show BALs (e.g.
Monier, Turnshek, and Lupie 1998).  HS 1216$+$5032
with $\Delta\theta = 9\arcsec$ is a
binary QSO with one BALQSO and one non--BALQSO (Hagen et al.~1996).

Here we report the serendipitous
discovery that LBQS 0103$-$2753 is a binary quasar with a projected
separation of 0\arcsecpt30 on the sky.
The two components show large differences in their emission-line spectra,
and only one has BALs.
LBQS 0103$-$2753 is one of eight BALQSOs observed in a program 
of HST/STIS and ground based spectrophotometry of low
and medium redshift BALQSOs for studies of chemical
abundances.
The selection criteria were the
presence of BALs (both high and low ionization BALQSOs
were candidates), redshift in the range
0.4 $<$ $z$ $<$ 1.8, and high apparent brightness.
LBQS 0103$-$2753, $z_e$ = 0.848, was chosen based on
its apparent magnitude, $B_J$ = 18.1, and its LBQS survey spectrum
showing broad, deep Mg II $\lambda$2798 absorption
around $z_a$ = 0.769 (Morris et al. 1991).

\section{OBSERVATIONS}

Observations  of LBQS 0103$-$2753 were obtained
on 1999 July 17--18 (UT). 
Figure 1 shows the STIS CCD acquisition image, which involved a
10s integration time in the F28X50LP mode 
(transmission from about 5000~\AA\ to 10,000~\AA, $\lambda_{pivot}$ =
7228.5~\AA) with a standard 100 by 100 subimage readout.
The image has total counts of 1113 and 356 DN for A and B
(formal errors 1.8\%, 4.1\%), respectively, using a 3 pixel radius
photometry aperture and a much larger annulus for sky.
The A/B flux ratio is 3.1, probably a slight underestimate
due to the wings of image A leaking into image B.
The centroids are
separated by $\Delta\theta$=0\arcsecpt295 $\pm$ 0.011.
The FWHMs of the images are 2.1 and 2.3 pixels for A and B, respectively,
consistent with point sources.
The position angle (PA) of the image $+y$ axis is
PA = $-$149.7, and the PA connecting A and B is $-$149.9$\deg$ or
+30.1$\deg$ $\pm$ 2.2$\deg$, with A north and east of B.


The QSO pair is almost perfectly aligned with the STIS slit,
itself aligned along the image +y axis.
A three step peak--up into
the 52X0.2 (0\arcsecpt2 wide) aperture was used for
the spectroscopic observations with
the STIS NUV--MAMA using the G230L grating.
Four orbits gave a total integration time of 10,907~s.
The spectra of the two QSOs are shown in Figure 2.
The spectra were extracted from the NUV--MAMA image using the
standard STSDAS/calstis routine ``x1d'' with an 11 pixel wide
extraction box for each QSO.  The measured separation of the two
spectra near the center of the NUV--MAMA image is 11.9 pixels
or 0\arcsecpt295, in agreement with the acquisition image.  Inspection
of plots perpendicular to the dispersion shows that the two spectra have
very little ($<$ 5\%) contamination from each other.
The nominal resolution at 2400~\AA\ is 2.1 pixels FWHM or 3.3~\AA\ for the
1.55~\AA\ per pixel sampling.  The useful wavelength range of the two
spectra is from 1650\AA\ (where the S/N $\sim$ 1) to 3145~\AA. The S/N in
the continuum around 2600~\AA\ is about 20 per pixel
for component A and 11 per pixel for B.


An optical spectrum of LBQS 0103$-$2753 (sum of both components)
was obtained obtained on 22 August 1998 UT using the CTIO Blanco
4m telescope.
The R--C spectrograph with the Blue Air Schmidt and Loral 3K CCD
with a 632 line mm$^{-1}$ grating was used to obtain a
3.0~\AA \ FWHM resolution spectrum covering 
3050~\AA \ to 5975~\AA .
The spectrum is shown in Figure 3, along with the LBQS spectrum
obtained on 26 August 1987 UT
by Morris et al. (1991).


\section{ANALYSIS}

At $z$ = 0.848, the projected separation of 0\arcsecpt30 amounts to 2.3 kpc
(\hnot = 70 \kms~Mpc$^{-1}$, $\Omega_M = 0.3$, $\Omega_{\Lambda} = 0.7$).
For comparison, the next smallest reliable 
binary quasar known, J1643$-$3156,
has an apparent separation of 15 kpc
(Brotherton et al.~1999).
Table 1 compares LBQS 0103$-$2753 to the next smallest binary quasars
and other BAL/non--BAL quasar pairs (also see lists by Mortlock et al. 1999 and
Kochanek et al. 1999).

The STIS spectra of LBQS 0103$-$2753 A and B show that the two
components of this QSO pair have very different emission,
continuum, and intrinsic absorption characteristics.
Component A shows strong BALs of 
C IV $\lambda\lambda$1548,1551 ($\sim 6,500$ to 27,000 \kms), Si IV $\lambda$1400,
N V $\lambda$1240, and O VI $\lambda$1034.
The emission line shapes, equivalent widths, and
intensity ratios are different between components A and B.

The two components of other binary QSOs agree in emission-line redshift within
about 600~\kms \ (Kochanek et al. 1999).  For LBQS 0103$-$2753
the emission redshifts, $z_A$ = 0.834 and $z_B$ = 0.858,
differ by 3900 \kms.
The centroids of the C IV $\lambda$1549
emission line profiles above 50\% of peak flux level
were used to measure the redshifts.
The formal one $\sigma$ errors in redshift for components
A and B, in velocity units,
are 370 \kms\ and 210 \kms\, respectively; and the
systematic errors, estimated from different choices for the continuum,
are $\pm 1000$ \kms \ and $\pm 400$ \kms.
For component B, the Ly $\alpha$ line gives the same redshift as C IV.
For component A, the Ly $\alpha$ and N~V
$\lambda$1240 emission lines are blended with an unknown ratio
of line strengths and an uncertain profile because the
N~V BAL truncates the feature on the blue side.
The Si IV and O IV] $\lambda$1400 and O VI $\lambda$1034
emission lines in component A,
with positions marked in Figure 2 based
on $z_A$ = 0.834, are probably present but weak and the
wavelengths are uncertain because
the shapes of the overall features are dominated by the adjacent BALs.
The redshift of component A depends entirely on the measurement
of the C IV line.
However, C~IV emission lines are often
blueshifted from the systemic redshift
(e.g. Espey et al. 1989); and weak low contrast lines, like
the  C~IV emission in A, tend to be shifted the most
(Brotherton et al. 1994).
If there was a
large blueshift from the systemic value for A 
and little blueshift from the systemic value for
B, the systemic redshifts could be close.  Further observations
are needed to determine the systemic redshift difference between A and B.

The optical spectrum of LBQS 0103$-$2753 (the summed components shown in
Fig. 3) has undergone a large change in 11 years in the observed
frame or 6 years in the QSO frame.
The Mg II $\lambda$2798 BAL, which is the strongest of the low
ionization BALs (Voit et al. 1993), has changed from about 50\% deep in 1987
to less than 10\% deep in 1998.
The spectrum of LBQS 0103$-$2753 obtained in August, 1998, matches the
average non--BALQSO spectrum (Weymann et al. 1991) to within 3\% between $\lambda
2600$ and the base of Mg II. Also the continuum has apparently become bluer from
1987 to 1998. Low ionization BALQSOs tend to have redder spectra than BALQSOs
without low ionization lines (Sprayberry and Foltz 1992).
The STIS spectrum of LBQS 0103$-$2753, obtained about 11 months after
the optical spectrum of
August, 1998, shows very strong high ionization BALs, but
no apparent low ionization BALs like C II $\lambda$1335.
In 6 years in the QSO frame, LBQS 0103$-$2753 A
has changed from a low ionization BALQSO to a high ionization BALQSO.
This is a large change compared with typical
BALQSO variability (Barlow 1993).
The unusual BAL variability was noted (Junkkarinen, Cohen, and Hamann 1999)
before the STIS observations
were obtained that show LBQS 0103$-$2753 is a binary QSO.

If we assume power-law continua $F_\lambda = F_0
(\lambda/\lambda_0)^p$ for each component, with $\lambda_0 = 5700$~\AA, the flux
ratios of A and B at
$\lambda3000$ (STIS) and $\lambda 7230$ (image) give $F_A/F_B =
3.0(\lambda/\lambda_0)^{0.2}$, so that B is slightly bluer than A.  Using the CTIO
fluxes at $\lambda$4150 and
$\lambda5700$, we find $p_A, p_B$ = -0.66,
-0.86 and $F_{0A}, F_{0B} = 2.4, 0.81$ 
in units of $10^{-16}~\ergpspcmpa$.  Extrapolating the observed spectra to
5500(1+z)~\AA, we find that components A, B have absolute magnitudes 
$M_V = -26.0, -24.6$, again for
$H_o = 70~\rm  km ~s^{-1}~ Mpc^{-1}$, $\Omega_m = 0.3$, and $\Omega_{\Lambda} =
0.7$.

\section{DISCUSSION}

The odds of finding two unrelated QSOs so close in direction and redshift are
negligible. The very different ultraviolet spectra of LBQS 0103$-$2753A,B show
that this is a true binary, not a lensed QSO.  Variability,
microlensing, and differential reddening might give different spectra or colors
for components of a lensed QSO.  However, for our pair the
expected time delay for a lens is $\sim$10 days, and luminous QSOs rarely show
significant variability over such times.  Unequal reddening cannot explain the
different absorption and emission lines of LBQS 0103$-$2753A,B (see Lopez
\etal\ 2000 for a related discussion).

What is the rate of occurrence of such close binary QSOs?  The
\hst\ snapshot survey for gravitationally lensed QSOs observed 498 QSOs with $z
> 1$ and $M_V < -25.5$ (\hnot=100, \qnot = 0.5; Maoz \etal\ 1993, and
references therein). No binary QSOs with separations less than 1\arcsecpt0 were
found. Our pair slightly misses the redshift cutoff for the
snapshot survey, but the angular diameter distance is not a strong function of
redshift above z = 0.86  (see Hogg 1999, and references therein).  LBQS
0103$-$2753 also misses the absolute magnitude cutoff by $\sim0.6$ magn.  However,
with regard to angular separation and magnitude
difference, the snapshot survey provides some guidance.
At a separation of
0\arcsecpt 3, the snapshot survey could detect pairs with a magnitude difference
$\le 1.5$ at a PA of 0\deg \ relative to the image trail, the least
favorable PA for detection (Bahcall \etal\ 1992).
A pair with the spacing and magnitude difference ($\Delta \rm m_V =
1.2$) of LBQS 0103$-$2753 should have been detected at any PA.
If the snapshot
survey accounts for most of the opportunities to detect such a pair to date,
the example of LBQS 0103$-$2753 is consistent with  a rate of roughly 1/500 for
separations $\sim$0\arcsecpt3. Among the $\sim10^4$ known
QSOs, there likely are of order 10 unrecognized binaries in the 0\arcsecpt3 range.

Is LBQS 0103$-$2753 likely to be a chance projection of a wider binary?
The probability of a QSO
having a binary companion at 3 to 10 arcseconds is $\sim10^{-2.7}$
(Djorgovski 1991; Kochanek \etal\ 1999, and references therein).  The cases
listed by Kochanek
\etal\ have a median separation of $\sim 5\arcsec$ (40~kpc).  The
probability that a pair with this physical spacing will have a projected 
spacing of 0\arcsecpt3 or less, because it is oriented nearly along the line of
sight, is (coincidentally)
$\sim10^{-2.7}$.  The probability that a randomly selected QSO will be such a
pair, and that it will be so oriented, is then $10^{-2.7}\times10^{-2.7} =
10^{-5.4}$.  Then, in a sample of 500 QSOs, the probability of finding even one
pair at 0\arcsecpt3 is only $\sim10^{-2.7}$.  This suggests that 
LBQS 0103$-$2753 is not a chance projection of a $\sim40$~kpc binary, but rather
has a true spacing not much greater than 2 kpc.

Close binary QSOs presumably represent mergers of galaxies in which the
nuclei in both galaxies are currently active.  Observations of nearby galaxies
show that the nuclei of large galaxies typically contain supermassive black
holes (\eg, Kormendy \& Richstone 1995). The merger will likely produce a
supermassive binary black hole (Begelman, Blandford, \& Rees 1980). Simulations
of colliding disk galaxies by Barnes and Hernquist (1996)
show little perturbation of the incoming galaxies until the first
pericenter passage.  However, gas 
accumulates in the nucleus of each galaxy as the two orbit away from each
other.  The time to orbit to
an apocenter $~\sim40$~kpc and fall back is $\sim 10^{8.5}~\rm
yr$ for galactic masses relevant here (see below).  Subsequent orbital loops are
much smaller, and the central regions of the galaxies merge quickly, entailing a
massive concentration of gas in the nucleus. We speculate that the observed
3\arcsec\ to 10\arcsec\ binary QSOs mostly represent galaxy pairs undergoing the
loop following the first close encounter, and that  LBQS 0103$-$2753 is in a later
stage when the nuclei of the two galaxies are coalescing.

Is the incidence of $0\arcsecpt3$ binaries consistent with this
scenario? The time for the orbit to decay from radius
$R$ to the center by dynamical friction is 
$
t_{df} \approx 
  (10^{8.6}~\yr) (R/2\rm kpc)^2 \,M_8^{-1}\,\sigma_{200},
$
(Binney \& Tremaine 1987; Mortlock \etal\ 1999, and references therein).  Here
$M_8$ is the
mass of the orbiting black hole (units $M/10^8\msun$),  assumed to be the
smaller hole on a circular orbit through the stellar background of the larger
galaxy (velocity dispersion 
$\sigma_{200} \equiv \sigma/200~\kmps$).  
Let us assume that each QSO is shining at one-third of the Eddington limit  
(see Kaspi \etal\ 2000) and take $L_{bol} \approx
8.3\lambda L_{\lambda}(3000~\rm\AA)$  following Laor (1998). Then $L_{bol}
\approx 10^{46.6}, 10^{46.1}$ for A, B; and the black hole masses are $M_A, M_B
\approx 10^{9.0}, 10^{8.5}$.   The recently discovered correlation 
of black hole mass with bulge velocity
dispersion, $M_{BH} = (10^{8.1}~\msun)\,\sigma_{200}^{3.75}$ (Gebhardt
\etal\ 2000, Ferrarese and Merritt 2000) gives
$\sigma_A, \sigma_B = 350, 260~\kmps$. The total bulge masses should be
$~\sim500 M_{\rm BH}$ (Kormendy \& Richstone 1995). Tidal radius considerations
suggest that the dense core of the stellar nucleus of galaxy B remains intact
around black hole B, with a total mass $\sim10^{10}~\msun$. This shortens the
dynamical friction time, and the above formula then suggests that the orbit decays
on the dynamical timescale of
$~\sim10^7$ years.  The discovery of LBQS 0103$-$2753 then indicates that
the probability of both nuclei being active as QSOs during the 2 kpc stage of a
merger is $\sim10^{1.5}$ times higher than during the 40 kpc phase.
Simulations of galactic mergers
including supermassive black holes could help to clarify these timescales.

Ultraluminous infrared galaxies (ULIGS, $L_{ir} >
10^{11}~\lsun$) typically involve mergers, the mean nuclear spacing being
$\sim2$ kpc (Sanders and Mirabel 1996; Genzel and Cesarsky 2000).  For high
luminosities, the predominant power source appears to be AGN activity that is
heavily obscured by a large mass of nuclear gas and dust.   From this perspective, 
an important aspect of LBQS 0103$-$2753 is the
lack of heavy extinction, perhaps because the high QSO
luminosity has dispersed the nuclear gas.  Infrared observations
of LBQS 0103$-$2753 would be most interesting, as would observations that could
detect a large mass of atomic or molecular gas expanding away from the nucleus.

The discovery of LBQS 0103$-$2753 suggests that binaries with separations of
several kpc occur with a frequency of order one or two per thousand QSOs.  The
detection of just a few more cases would provide a better estimate of this
rate, providing an important constraint on galactic mergers and the fueling
of active galactic nuclei. 

\acknowledgments

We thank R. Blandford, K. Gebhardt, L. Hernquist, J. Kormendy, C. Kochanek, 
C. Mihos, S. Phinney, D.
Richstone,  and R. Weymann for helpful discussions.  We also thank R. Weymann for
providing in digital form the LBQS 0103$-$2753 spectrum from the LBQS survey, and
K. Gebhardt for the use of a computer routine.
G.A.S. gratefully acknowledges the hospitality of Lick Observatory and Stanford
University during summer 2000.
Support for this work was provided by NASA through grants
GO-07359, GO-07359.02 from the Space Telescope Science Institute,
which is operated by the Association of Universities for Research in
Astronomy under NASA contract NAS5-26555.

\clearpage

\clearpage
\begin{deluxetable}{llrrcrll}
\tablecolumns{8}
\footnotesize
\tablecaption{Binary Quasars and Binary Quasar Candidates\label{tbl-1}}
\tablewidth{0pt}
\tablehead{
\colhead{Name     } & \colhead{$z_e$  } & \colhead{$\Delta\theta$} &
\colhead{$R$\tablenotemark{a} } & \colhead{$m$    } & \colhead{$\Delta m$    } &
\colhead{comment  } & \colhead{REF.\tablenotemark{c}   }\\
\colhead{}          & \colhead{}        & \colhead{\arcsec     } &
\colhead{kpc      } & \colhead{brighter} & \colhead{}              &
\colhead{}          & \colhead{}      
}
\startdata
\multicolumn{8}{c}{Close Pairs } \\
\tableline
LBQS 0103$-$2753  & 0.848 & 0.3 & 2.3 & 18.2 & 1.2: & BAL/non--BAL pair  & 1  \\
J1643$+$3156      & 0.586 & 2.3 & 15  & 18.4 & 0.8 & radio--loud/radio--quiet
pair&
 2 \\
CTQ 839           & 2.24  & 2.1 & 17  & 18.2 & 1.9 & radio--quiet/radio--quiet
pair   &  3 \\
\tableline
\multicolumn{8}{c}{} \\
\multicolumn{8}{c}{Other BAL/non--BAL pairs }\\
\tableline
LBQS 2153$-$2056  & 1.85  & 7.8 & 66  & 17.9 & 3.4 & BAL/non--BAL pair  &  4 \\
HS 1216$+$5032    & 1.45  & 9.1 & 77  & 17.2 & 1.8 & BAL/non--BAL pair  &  5,6
\\ Q1343$+$2640\tablenotemark{b}      & 2.03  & 9.5 & 79  & 20.2 & 0.1 & weak
BAL/non--BAL pair & 7\\
\enddata
\tablenotetext{a}{The projected separation $R$ is calculated using
$H_o = 70~\rm km ~s^{-1} ~Mpc^{-1}, \Omega_m = 0.3, \Omega_{\Lambda} = 0.7$.}
\tablenotetext{b}{Q1343$+$2640 is also a radio--loud radio--quiet
pair (Kochanek, Falco, and Mu\~noz 1999)}
\tablenotetext{c}{REFERENCES.---(1) this paper. (2)
Brotherton et al.~1999.  (3) Morgan et al.~2000
(4) Hewett et al. 1998. (5) Hagen et al. 1996.
(6) Lopez, Hagen, and Reimers 2000.
(7) Crotts et al. 1994.}
\end{deluxetable}

\clearpage
\vskip3.0in
\begin{figure}
\plotone{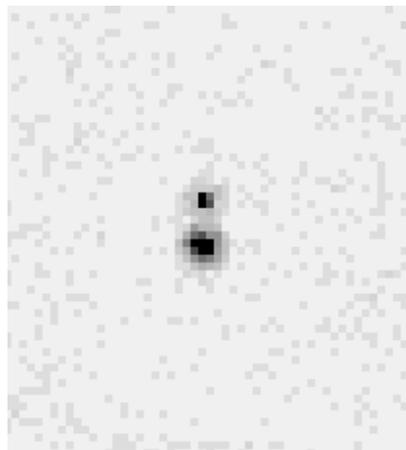}
\figcaption[fig1.ps]{Acquisition image of LBQS 0103$-$2753 obtained with
the STIS CCD in the 28X50LP mode.
The brighter object is A, the BALQSO, and the fainter is B.
The PA of the +$y$ axis in this picture is at $-$149.7$\deg$
measured East from North.
\label{fig1}}
\end{figure}

\clearpage
\vskip3.0in
\begin{figure}
\plotone{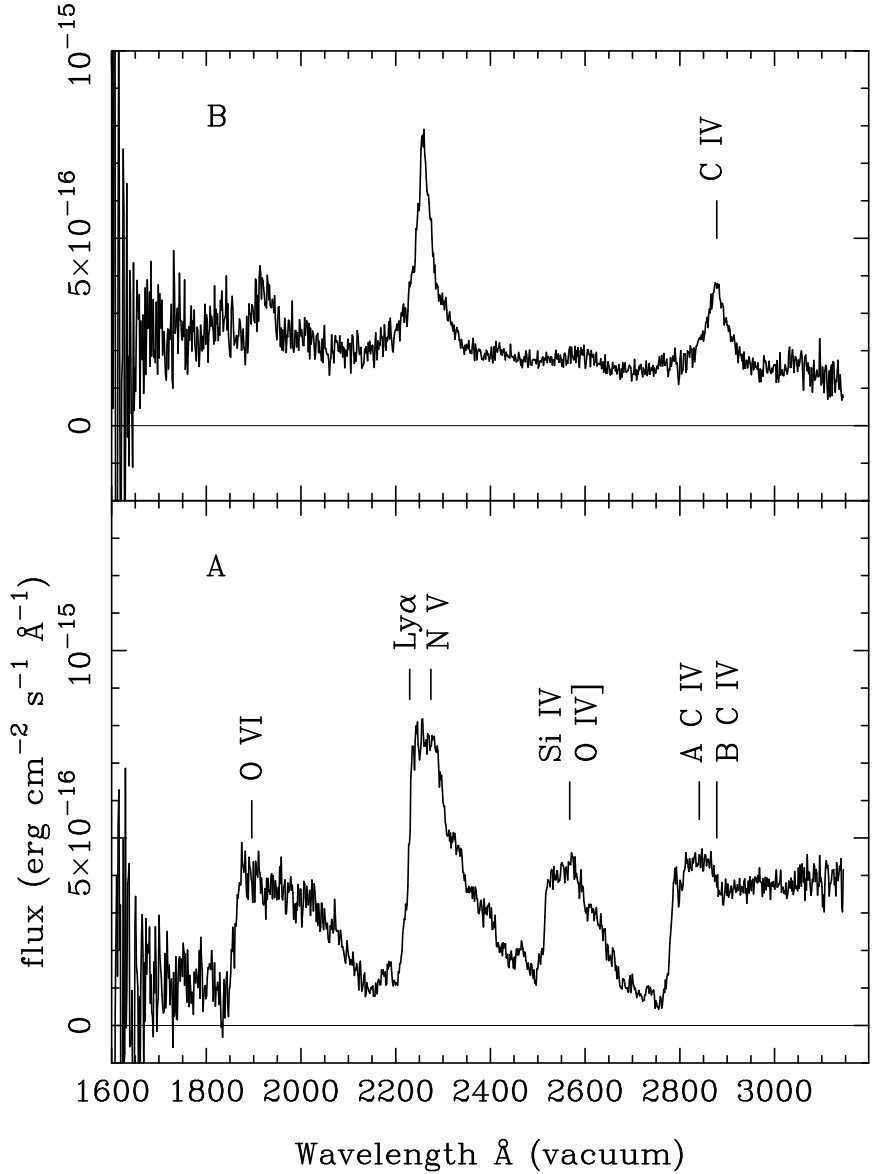}
\figcaption[fig2.ps]{Spectra of LBQS 0103$-$2753 A and B
obtained with HST/STIS using the NUV--MAMA and G230L grating.
The upper panel is
component B and the lower panel is component A.
The measured C IV $\lambda$ 1549 emission-line wavelengths for components A and
B are both shown with vertical marks above component A's CIV
$\lambda$ 1549 emission line.
\label{fig2}}
\end{figure}

\clearpage
\vskip3.0in
\begin{figure}
\plotone{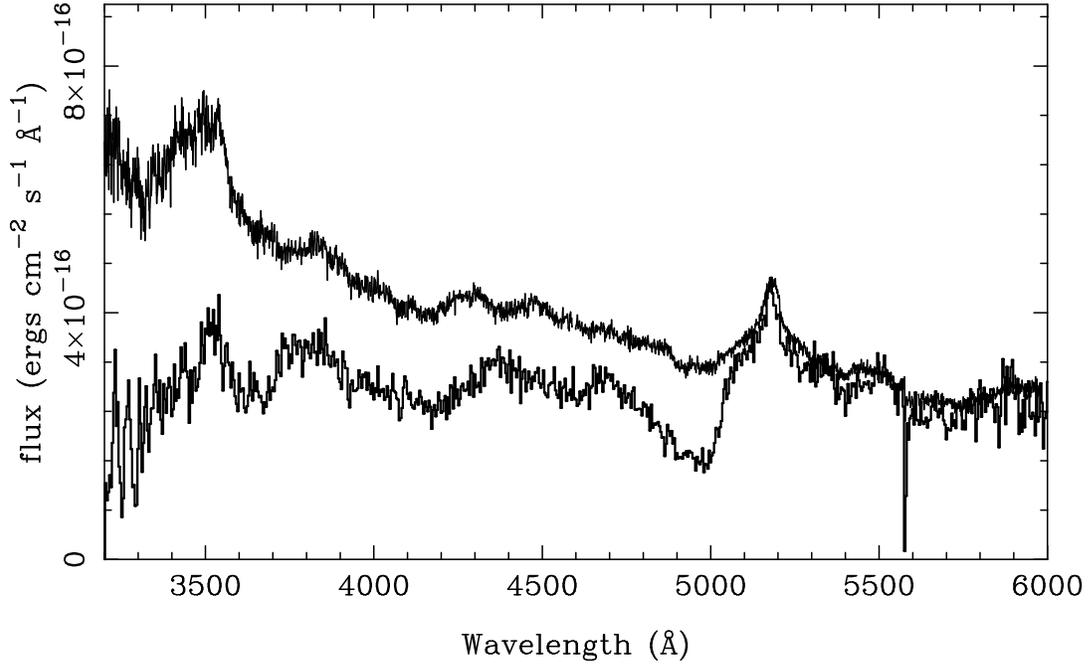}
\figcaption[fig3.ps]{Spectra of LBQS 0103$-$2753 A plus B
(the sum of the images).  The upper spectrum is from CTIO
on 22 August 1998 UT, the lower spectrum, at somewhat lower
resolution and S/N, is from the LBQS survey obtained on
26 August 1987 UT (Morris et al. 1991).  The strong absorption
around 4950~\AA\ in the LBQS spectrum is a Mg II $\lambda$2798
BAL.   The LBQS spectrum has been scaled to make the
spectra match at the red end.
\label{fig3}}
\end{figure}

\end{document}